\documentclass[preprint,showpacs,preprintnumbers,amsmath,amssymb]{revtex4}

\def\OO{{\cal O}}
\def\NN{{\cal N}}

\def\none{$\NN=1$}

\def\ntwo{$\NN=2$}
\def\nfour{$\NN=4$}

\def\ZZ{{\bf Z}}

\def\tr{{\rm tr}}

\def\bel{\begin{equation}\label}
\def\ee{\end{equation}}

\newcommand\vev[1]{\langle{#1}\rangle}

\def\mod{\ {\rm mod}\ }

\def\Lgr{{\cal L}}
\def\LUV{\Lambda_{UV}}
\def\Lco{\Lambda_{co}}
\def\mass{{\cal K}}
\def\mn{{\mu\nu}}
\def\lsim{\alt}
\def\gsim{\agt}

\begin{document}
\preprint{UW/PT 03-16, hep-th/0309122
}
\date{\today}
\title{
Non-Supersymmetric Theories with Light Scalar Fields and Large Hierarchies}
\author{
Matthew J. Strassler}\affiliation{Department of Physics, 
University of Washington, P.O.~Box 351560, Seattle WA 98195-1560 USA}

\begin{abstract}
Various nonsupersymmetric theories at large but finite $N$ are argued
to permit light scalars and large hierarchies without fine-tuning.  In
a dual string description, the hierarchy results from competition
between classical and quantum effects.  In some cases the flow may end
when a string mode becomes tachyonic and condenses, thereby realizing
a quantum-mechanically stable Randall-Sundrum hierarchy
scenario. Among possible applications, it is suggested that lattice
simulation of \nfour\ Yang-Mills at large 't Hooft coupling may be
easier than expected, and that supersymmetry may naturally be an
approximate symmetry of our world.  (This letter is a writeup of work
presented at Aspen in summer 2002.)
\end{abstract}

\maketitle

Naively, nonsupersymmetric field theories with light scalar fields can
only be obtained through fine tuning.  A relevant mass operator which
is a singlet under all global symmetries can always be obtained by
replacing every $\partial_\mu$ in the scalar kinetic terms with a mass
parameter $m$.  In complete generality, any global-singlet relevant
operator (GSRO) naturally has a coefficient of order a power of some
cutoff $\Lambda_{UV}$; thus $m^2\sim\LUV^2$ unless fine-tuning is
applied.  Indeed, the Higgs boson mass is the only GSRO in the
standard model of particle physics, and is the source of the hierarchy
problem.

There are well-known ways to evade this constraint.  Supersymmetry
(SUSY) relates scalar and fermion mass operators; the latter can be
forbidden by a chiral symmetry.  For Nambu-Goldstone bosons, the shift
symmetry $\delta\phi =$ {\it constant} eliminates the mass operator.
Conformal symmetry forbids all dimensionful coefficients.  However,
these symmetries cannot be realized as accidental symmetries in
theories with GSROs; if they are not exact, a mass of order
$c\Lambda_{UV}^2$, where $c$ is at best an algebraically-small
parameter, will be generated.

Another recent example is similar.  The planar graphs of a SUSY gauge
theory are shared by its ``orbifolds'' \cite{orbifold}, obtained via a
projection operator that eliminates certain fields of the SUSY gauge
theory.  At $N=\infty$, then, the orbifold theories have $m=0$.
However, nonplanar graphs give $\delta m^2\sim \LUV^2/N^p$ where
$p\geq 1$.  Exponential suppression of $m$ requires exponentially
large $N$.

All of these ideas have been suggested for solving the standard model
hierarchy problem.  The SUSY story is well-known; the
Higgs-as-pseudo-Goldstone-boson has recently been made to work
successfully in deconstruction-motivated models \cite{decon}; and
conformal symmetry has been suggested by Bardeen \cite{bardeen} and in
the context of orbifold models by Frampton and Vafa \cite{fv}.  Any
new ideas might also be useful for model building.

In this article I suggest a novel mechanism which allows light
scalars. \footnote{Many readers will have seen
versions of this talk, given in 2001 and 2002.  The early 2001
versions were not quite correct, due to my imperfect understanding of
the role of double-trace operators.  Even now, the subtleties involved
with these operators and the crucial unknown sign of a beta function
still leave some important questions unanswered.  I would like to
thank Markus Luty, who though conversations surrounding his very
interesting and highly recommended paper \cite{Markus} with Goh and
Ng, has motivated me to publish this incomplete assessment of these
issues.  Clearly there is more to say on this subject.}\footnote{Since the time
that this talk was given, a number of other important related papers
have appeared \cite{Troost:2003ig,Gubser:2002zh,Gubser:2002vv}.}  
The above arguments rely on perturbative intuition, for
which $m^2$ has positive mass dimension.  However, if the mass
operator has a large anomalous dimension, it may actually be
irrelevant; in this case $m^2$ may be present, but it runs to zero
(faster than the Wilsonian renormalization scale) and its infrared
effects are suppressed.  Using AdS/CFT duality \cite{malda}, theories
can be constructed which have no GSROs whatsoever; correspondingly,
the dual string theories contain no gauge-singlet tachyons.  In some
cases these models allow (and may generate) exponentially small scalar
masses and large hierarchies.  (In the string description, this
happens through a balancing of classical ($N^0)$ and quantum
$(N^{-2}$) effects.)  These non-SUSY theories naturally realize
Randall-Sundrum-like hierarchies \cite{RS} without quantum
inconsistencies or instabilities.

Consider $SU(N)$ $d=4$ \nfour\ SUSY YM, with gauge coupling $g_{YM}$;
define $\alpha = g_{YM}^2/4\pi $, and the 't Hooft coupling
$\eta\equiv \alpha N$.  Its dual is IIB string theory on $AdS_5\times
S^5$, with string coupling $g_s\equiv \alpha$.  By hand, add a UV
cutoff at which SUSY is completely broken {\it but $SO(6)$ is
preserved.}  The effective action at $\LUV$ includes all possible
$SO(6)$-singlet operators with coefficients of a natural size.
Classically, this theory has dimension-2 scalar bilinears $\mass =
N^{-1}\tr \phi^k\phi^k$ and $\OO_2^{ij}=N^{-1}\tr
\phi^i\phi^j-{1\over 6}\delta^{ij}\mass$, in the ${\bf 1}$ and
${\bf 20'}$ of $SO(6)$.  The former is a GSRO which we cannot forbid by
any symmetry.  There are no $SO(6)$-singlet dimension-3 operators.  At
dimension 4 there are several global-singlet marginal operators
(GSMOs): the single-trace operators which appear in the Lagrangian,
and the double-trace operators $\mass^2$ and $\OO_2^2\equiv
\sum_{ij}\OO_2^{ij}\OO_2^{ij}$.

For $\eta\ll1$, the dynamics is entirely determined by the operator
$\mass$, which is allowed by all symmetries and will rapidly dominate
the infrared.  The scalars are massive, but the
four  adjoint fermions remain massless.  The IR
behavior of this theory is unknown.

The situation at $\eta\gg 1$ is different.  Take $N$ {\it large but
finite}, of order, say, 10-100, and choose $\eta$ {\it large but finite}, 
of order 5-50, at the cutoff $\LUV$.
The operators of the gauge theory correspond to modes of supergravity
(SUGRA) fields on $AdS_5$.  All modes with dimensions of order 1 are
classified in \cite{KRN}; {\it there are no GSROs!}  In fact $\mass$
corresponds to a massive string mode (see {\it e.g.} \cite{arut}), and
has dimension of order $(4\pi\eta)^{1/4}$ \cite{malda}, greater
than four.  Consequently the effect of $\mass$ {\it decreases} in the
IR, and the scalars remain light despite the lack of SUSY.

Absent a GSRO, the theory is governed by two GSMOs: 
the double trace operator 
$\OO_2^2$ and the \nfour\ Lagrangian itself, $\OO_4=N^{-1}\tr(F_\mn F^\mn+\dots)$. 
With the $\theta$ angle zero for simplicity, the Lagrangian is
$$
\Lgr = {N\over 2g^2} \left[\OO_4 - {h^2 N^2\over 4g^2N}\OO_2^2 
 \right]
$$
plus boundary terms and terms that vanish on the equations of motion.
The expansion parameters for perturbation theory are $\eta =
g^2N/4\pi$ and $\xi\equiv h^2N^2/4\pi$.  The $\xi\ll 1\ll \eta$ region
is controlled by conformal perturbation theory around \nfour\ SYM.
The operator $\OO_2^2$ is marginally {\it relevant}, with
${\rm dim}(\OO_2^2) -4 = - ({16/N^2}) $ for $\eta\gg 1$  \cite{arut} 
and $-(5/ 2\pi) (\eta/ N^2)$ for small $\eta$ \cite{arutrecent}.  [That the correction is
of order $1/N^2$ follows from large-$N$ factorization of
$\vev{\OO_2^2(x)\OO_2^2(0)}$ and nonrenormalization of $\vev{\OO_2^{ij}(x)\OO_2^{kl}(0)}$.  
In SUGRA $\OO_2^2$ corresponds to a
non-BPS particle-antiparticle state, with a negative binding energy
--- hence the sign.]  

Note we may now (if we wish) take $\LUV\to\infty$,
defining without fine tuning a renormalizable 
theory which is \nfour\ SUSY in the UV with SUSY broken by
$h\neq 0$.  The breaking is soft due to the absence of GSROs.  
For now we will keep $\LUV$ finite (since in the real world gravity
will always add additional physics) while keeping in mind that
this limit may always be taken.

I will now address the following issues.  First, I will show 
the beta functions for $\eta$ and $\xi$ are at most of order $1/N^4$.
Next, I will argue the theory has no Coleman-Weinberg instabilities 
except possibly at exponentially small energy scales.  Finally, although
I will not determine the sign of $\beta_\eta$, I will show that for
either sign interesting physics, and a large or infinite hierarchy, results.

{\it Beta Functions I:}
The dimensionless coupling which determines the validity of
perturbation theory is $h\mu^{-\dim(h)}$.  For this
reason we redefine $\xi(\mu) = h^2(\mu)\mu^{-2\dim(h)}N^2/4\pi$. This
coupling blows up, classically, at 
$\Lambda_\xi \sim \xi_0^{N^2/16}\LUV\ll\LUV $, where $\xi_0$ is
$\xi(\mu=\LUV)$.  But in fact $\xi$ never becomes large. At
order $\xi^2$ the $\beta$ function receives a large {\it positive}
correction \cite{ewdbltr} 
\bel{betaxi} \beta_{\xi} =
\xi\left(-{16\over N^2} +  {\xi\over 4\pi}\right) \ .  \ee
[
The order-$\xi$ correction to $\dim\OO_2^2$ is
$$\propto \int d^4z\vev{\OO_2^2(x)\OO_2^2(0)\OO_2^2(z)} \ .$$ 
This factorizes, so it is
order $N^{0}$; its sign is  as in $\phi^4$ theory.]
Thus the flow drives $\xi$ to  $\sim 64\pi/ N^2$ and $\beta_\xi\lsim 100/N^{4}$.  

Meanwhile, $\xi=0$ implies $\beta_\eta=0$ by \nfour, and the
leading $\xi$-dependence involves connected graphs, so
$$
\beta_\eta = C(\eta){\xi\over N^2} .
$$
where $C(\eta)$ is an unknown function of order $N^0$.  [In SUGRA,
the $N^{-2}$ can also be obtained:
the addition of $\OO_2^2$ to the action \cite{absdbltr}
involves a change of boundary conditions for the mode $U^{ij}$ corresponding to 
$\OO_2^{ij}$ \cite{ewdbltr,berkdbltr}
but for $\vev{\OO_2^{ij}}=0$ this can have no classical effect on the dilaton.]
Since $\xi$ itself is order $1/N^2$, $\beta_\eta\sim 1/N^4$. Thus
all beta functions are small and no appreciable change of couplings occurs
above scales of order $e^{-cN^4}\LUV$, where $c$ is $N$-independent but could
be $\lsim 10^{-2}$.

{\it Stability:}  Could there be a Coleman-Weinberg instability?  The
\nfour\ SUSY YM has a moduli space; classically, most is lifted by the
$h^2\OO_2^2$ perturbation, but some flat directions in the potential
energy remain. Could quantum corrections make the
origin unstable? The approximate conformal invariance at the origin
of moduli space
implies any effective potential whose overall scale is $v$ scales as
$v^4$ near the origin.  Along any classically-flat direction, the
$SU(N)$ gauge group is broken to a product group which contains
decoupled $U(1)$ factors, whose scalars have an $\OO_2^2$ interaction.
These factors are at small 't Hooft coupling and can be studied in
perturbation theory.  As with any $\phi^4$ interaction, a {\it
positive} effective mass-squared $m_{eff}^2\phi^2$ will be generated;
but the relevant cutoff is $v$, so altogether this represents a
positive $v^4$ term in the potential.  This suggests the theory at
$v=0$ is stable.  Of course the argument receives corrections near
$v=\LUV$ (but in this regime, couplings of irrelevant operators can be
chosen so as to avoid any minima in the energy) and from the non-zero
running there may be effects at exponentially small $v$ (but any
minima in this regime would be interesting rather than problematic.)

{\it Beta Functions II.}
Let us assume the above argument is correct and address the issue of $\beta_\eta$.
Unfortunately the computation of this quantity, or even the sign,
is problematic, and will not be attempted here.  Instead,
let us note that both signs are interesting.

\underline{$\beta_\eta<0$:}  In the IR $\eta$ becomes large; classical string
theory breaks down.  Since \nfour\ has $g_s \to 1/g_s$ duality, the
beta function must change sign at large $\eta$, suggesting \cite{fv}
the presence of at least one IR-stable conformal fixed point.
This fixed point will have massless scalars (but no
moduli space, see above). It is probably
non-SUSY, although there is also the possibility of accidental  \nfour\ SUSY.
If there is no such fixed
point, then something else interesting must happen when $\eta\sim N$, 
leaving the scalars exponentially light and/or with exponentially small
vevs. 

\underline{$\beta_\eta>0$:}  Now $\eta$ decreases in the IR.
Let us call the scale where $\eta\sim 1$ the cross-over scale
$\Lco$.  
What can happen at this scale? Since $\dim \OO_2^2 \to 4$
as $\eta\to 0$, the coupling $\xi$ remains small.  There might
be a fixed point for $\eta\sim 1$, but this seems unlikely,
because in this region the operator $\mass$ is a GSRO, making
accidental conformal invariance impossible.  Instead,
$\mass$ acts as a``dangerous irrelevant'' operator; its
coefficient $m^2$ shrinks for $\mu>\Lco$ 
but grows for $\mu<\Lco$, becoming large
and dominating the IR physics.  Note $m^2$,
although
multiplicatively renormalized in the SUSY theory, is
{\it additively} renormalized when $\xi\neq 0$, and is not driven
exactly toward zero.  Therefore, at $\Lco$, $|m^2|\gsim \xi \Lco^2$; it
cannot be smaller.  (The sign of the additive renormalization is
presumably positive, as in $\phi^4$ theory, 
though not strictly calculable at $\eta\sim 1$.)
Thus, the mass does not generate a scale {\it exponentially} small 
compared to $\Lco$: the mechanism of Bardeen \cite{bardeen} --- in
which scalar masses are protected by the scale invariance of the
theory --- is not quite realized.  Rather,  the {\it irrelevance} of
$\mass$ has led to a scale exponentially smaller than $\LUV$ and
of order $\Lco/N$.

In both cases, any expectation values, dimensionful coupling constants
and/or scalar masses lie exponentially [in $N^4$] below $\LUV$, without
any fine tuning.  
Having established this result, I will make a few comments.
Thereafter, I will
discuss other models with similar properties, as well as variants
which more certainly have positive $\beta_\eta$.

{\it Dual string description:} In the AdS description (with
radial coordinate $r\propto \mu$) no GSROs means no gauge-singlet
tachyonic modes at large $r$.  Classically the only effect (away from
$r_{max}\propto\LUV$) of SUSY breaking is the altered boundary condition on the
mode $U^{ij}$ corresponding to $\OO_2^{ij}$.  However this boundary
condition affects the the dilaton $S$ at the quantum level, causing it
to vary ($dS/dr \propto \beta_\eta$)  along with the curvature radius.  
The absence of a Coleman-Weinberg instability
implies the force (classical plus one-loop)
between D3-branes in the presence of the new boundary condition is
attractive; thus the branes will remain in a single clump
and the nearly-AdS region
will be preserved.  If $\beta_\eta>0$ the string coupling and
curvature radius decrease until, at some small $r_{min}\propto\Lco$, the lightest
string state becomes tachyonic, and condenses at and/or inside this
region (possibly via both non-normalizable and normalizable modes.)
Thereafter the dynamics of the theory becomes difficult to guess, but
the space certainly cuts off near this point. If $\beta_\eta<0$ the
string coupling gradually increases as $r$ decreases;
when $\eta\sim N$, a fixed
point may be reached, or perhaps other dynamics sets in.  

{\it Model building:} With no IR fixed point, the theory generates a low {\it non-zero} scale
automatically; but even when there is a fixed point, it is easy to dynamically
generate a low non-zero scale at which
scalar masses would appear.  Simply couple this theory through $SO(6)$-preserving
irrelevant operators to a SUSY or non-SUSY small-$N$ sector which
spontaneously breaks $SO(6)$.  This would cause dynamical conformal
symmetry breaking at a yet lower scale in the large-$N$ sector.
In string language, the dynamics
of the other sector acts as a small source for an $SO(6)$ non-singlet tachyon,
causing it to condense at some small $r$.

{\it The cosmological constant:}  This theory does not have a naturally small
cosmological constant.  The unit operator is relevant, so any
violation of SUSY at the scale $\LUV$ will generate a cosmological
constant of order $\LUV^4$.  Even if the theory is SUSY at $\LUV$ (leaving 
$\Lambda_\xi$, the classical SUSY-breaking scale, finite) 
we will still get a large cosmological
constant of order $h^2\LUV^{4-2\dim(h)}\gg\Lambda_\xi$.

{\it Supersymmetry:} In this theory, SUSY may be badly broken
in the ultraviolet, restored up to $1/N^2$ corrections in an
intermediate regime, and broken badly in the infrared.  Could this be
the story in the real world?  Could SUSY be everywhere at best
approximate, yet still control the physics above the weak scale
because of the absence of GSROs?  To make a realistic model  
seems difficult; nonetheless the possibility is
thought-provoking.\footnote{This line of thought
has been considered every now and then over the years,
the earliest dating back to \cite{kaplan}.  A more serious
recent discussion, along the lines considered here, has
just appeared \cite{Markus}.}

{\it Lattice gauge theory:}  Simulating \nfour\ SUSY on the lattice
seems hopeless, requiring excessive fine-tuning.  But if $\eta\gg 1$,
the necessary tuning may be minimal; if enough of $SO(6)$ is
preserved, then there are no GSROs and few unwanted GSMOs, whose
couplings are naturally driven small.  How large must $N$ be for this to
be useful? Surely not 100, but is $N=10$ sufficient?  Although not
feasible at present, it is remarkable that the difficulties of
simulating \nfour\ at large $\eta$ --- and with it the predictions of
string theory --- might be be algebraic rather than exponential.
[Alternatively, one might remove the scalars and
replace them with local four-fermion couplings; the scalars might be
regenerated as in the Nambu--Jona-Lasinio model.]

{\bf More Examples:}
One may consider other theories which work similarly; in each of these
examples the arguments for energetic stability will have to be
revisited.

{\it Other nearly-supersymmetric theories:}  The \none\
SUSY-preserving ${\bf C}^3/\ZZ_3$ orbifold, which breaks $SO(6)\to
SU(3)\times U(1)$, works the same way; there are no GSROs and a
couple of GSMOs.  [The absence of a dimension-two GSRO follows from
the fact that the ${\bf 20'}$ has no singlets under $SO(6)\to
SU(3)\times U(1)$.)]  However, all \ntwo\ orbifolds of \nfour\ have a
GSRO, an operator with positive (negative) $m^2$ for vector-multiplet
(hypermultiplet) scalars.  ({\it E.g.}, the ${\bf C}^2/\ZZ_2$
orbifold preserves an $SO(4)\times SO(2)$ of $SO(6)$, under which
${\bf 20'}\to{\bf 9_0} \oplus \bf{4_1} \oplus \bf{4_{-1}}
\oplus \bf{1_2} \oplus \bf{1_{-2}} \oplus \bf{1_0}$; the last is the
GSRO.)  Most \none\ models fail for similar reasons.  In short, the
physics under discussion is not generic.

{\it Fully non-supersymmetric examples?}  More interesting would be a
non-SUSY orbifold, which would realize the program of Frampton and
Vafa \cite{fv}.  As noted earlier, scalars in these models naturally
have masses $\sim \LUV/N$ in perturbation theory, far too large to
explain the observed hierarchy.  For $\eta\gg1$, this does not apply
if an orbifold can be found with no GSROs.  However we must also
ensure that there be no disallowed tachyons --- instabilities which
would ensure that there is no unitary conformal field theory at all.
Unfortunately, most non-SUSY orbifolds have either fixed planes with
disallowed tachyons or GSROs; no problem-free example is known.

{\it Non-supersymmetric conifold:} A different approach is
provided by D3 branes at a conifold singularity.  Naively this appears
doomed: the operator $\OO_2^{ij}\sim\tr A^iB^j$ has dimension ${3\over
2}$, so its square is a GSRO.  However, as in \cite{kleb,ewdbltr}, one
may add auxiliary fields $Z_{ij}$ and SUSY-violating terms $\Delta\Lgr=
Z_{ij}\OO_2^{ij} + B Z_{ij}Z_{ij}$. The operator $(Z_{ij})^2$ has
dimension 5 and $B$ runs to zero in the IR, leaving $\OO_2^{ij}$
redundant.  In SUGRA, this is precisely the reversed quantization condition
on $U^{ij}$
discussed in \cite{ewdbltr}.  
The global $SU(2)\times SU(2)\times U(1)$ allows no GSROs
in this  theory; there are two-fermion/two-scalar GSMOs, and
the analysis proceeds as in our original example. 

{\bf Driving $\eta$ small:} The case $\beta_\eta>0$ is especially elegant; the theory 
breaks its own approximate conformal invariance
through a dangerous irrelevant operator,
without requiring an additional sector.  Here are some examples which
might realize this scenario even if the original example does not.
Again, in all these cases energetic stability must be
reconsidered.

{\it Addition of fermions:}  Added hypermultiplets (making the theory
non-asymptotically free, but recall $\Lambda_{UV}$ is finite)
would give $\beta_\eta>0$ but would add GSROs, as in all \ntwo\
theories.  However, addition of $N_f>1$ {\it fermions} in the ${\bf
N}\oplus {\bf\overline N}$ 
representation apparently introduces neither GSROs nor GSMOs,
while making $\beta_\eta\sim +N_f/N$. Unfortunately the scalar 
potential at the origin is probably not stable.

{\it Addition of gauge fields:} A non-SUSY gauging of $SO(3)\times
U(1)\subset SO(6)$ introduces no GSROs (though there are additional
GSMOs.)  The new gauge couplings $\hat g$ rapidly run small, so $\beta_\eta\sim
\hat g^2$ itself rapidly becomes small.
Can $\eta$ be driven into a region where $\mass$ is
relevant?  Unfortunately the
effect on $\eta$ is borderline, leaving the answer ambiguous.  
(More generally, many possible small-$N$ field theories could
be coupled to the large-$N$ theory while preserving part of
$SO(6)$; in some cases $\eta$ might be pushed from
large to small.)

{\it Duality cascade:}  Consider the theory of  $M>2$
fractional $D3$ branes at the conifold singularity \cite{klebnek,ikms}, 
with the relevant double-trace operator again removed by 
Lagrange multipliers $Z_{ij}$.  The symmetry group
$SU(2)\times SU(2)\times \ZZ_{2M}$ still forbids all GSROs.
A duality cascade ensues \cite{ikms}, reducing $N$ with $g_s$ fixed, and
thereby making $\beta_\eta\sim M/N$.  (The additional fields $Z_{ij}$
cannot ruin Seiberg duality; at small $\eta$ SUSY would be badly
broken, but without GSROs the SUSY breaking is extremely soft.)
If the theory is energetically stable, 
there are two possible endpoints for the
flow:

\underline{$g_sM\gg 1$:} in this case, the cascade ends, confinement occurs,
and $\ZZ_{2M}$ is spontaneously
broken at a scale $\LUV e^{-\eta_0/\eta_f^2}$,
where $\eta_0=\eta(\mu=\LUV)$, $\eta_f = g_s M$.
If $p\equiv N \mod M\neq 0$, then as emphasized in \cite{csaki} the continuous
global symmetries are spontaneously broken.  Since SUSY is explicitly
broken there will be only pseudoscalar Goldstone bosons without their
scalar SUSY partners; however, mass splittings, {\it etc.}, will be
small.  All of the physics can be reliably computed using SUGRA.

\underline{$g_sM\lsim 1$:} In this case, $\eta$ reaches 1, at a scale $\Lco$, before
$N\sim M$.  At this point, the scalar mass term becomes relevant and
the SUSY breaking becomes important; the cascade breaks down, as does
the SUGRA description.  The low-energy theory will be vaguely
standard-model-like; it will have scalar masses and apparent SUSY
breaking at this scale, a product gauge group with few colors (if $g_s$
is not too small), and some light fermions.  Precise details of the
physics appear difficult to compute but might be worth further study
(assuming the theory is energetically stable.)

In this and other models, spontaneous breaking of chiral symmetries is possible
or even automatic; are there applications to technicolor?  The
real problem in technicolor is not symmetry breaking but flavor, and it
is far from clear that the breaking dynamics can be coupled to
fermions in a realistic manner without introducing GSROs.  Moreover,
if the standard model is a spectator to the dynamics, then its beta
functions tend to be positive and order $N^2$ or $N$; this is a huge
obstacle to realism.  It seems more likely that the standard model
must be embedded inside a large-$N$ gauge theory for the dynamics
described here to be of particle-physics interest.  
But for any realistic application, it will
be necessary to find and understand a wider class of examples.
It is not easy to find theories without GSROs; as noted above there are
significant constraints.  

In summary, theories at large 't Hooft coupling can  violate 
standard notions of naturalness in interesting ways.  At this stage it
is impossible to guess whether  these phenomena are relevant
in the real world.  However, with new model-building 
tools come new ideas, and we may hope for
the best.

  I am grateful for the hospitality of the Institute for Advanced
Study, where the present version of this work was completed and to
A. Adams, V. Balasubramanian, R. Gopakumar, K. Intriligator,
I. Klebanov, A. Nelson, L. Randall, L. Rastelli, N. Seiberg, S. Sethi,
E. Silverstein, W. Skiba, R. Sundrum, and E. Witten for comments and
insights.  This work was supported by DOE grant DOE-FG02-95ER40893,
NSF grant PHY-0070928, and by the Alfred P. Sloan Foundation.


\begin{references}

\bibitem{orbifold}
M.~R.~Douglas and G.~W.~Moore,
arXiv:hep-th/9603167.
S.~Kachru and E.~Silverstein,
Phys.\ Rev.\ Lett.\  {\bf 80}, 4855 (1998)
[arXiv:hep-th/9802183],
%
A.~E.~Lawrence, N.~Nekrasov and C.~Vafa,
Nucl.\ Phys.\ B {\bf 533}, 199 (1998)
[arXiv:hep-th/9803015];
A.~Hanany, M.~J.~Strassler and A.~M.~Uranga,
JHEP {\bf 9806}, 011 (1998)
[arXiv:hep-th/9803086];
M.~Bershadsky, Z.~Kakushadze and C.~Vafa,
Nucl.\ Phys.\ B {\bf 523}, 59 (1998)
[arXiv:hep-th/9803076].
.

\bibitem{decon}
N.~Arkani-Hamed, A.~G.~Cohen and H.~Georgi,
Phys.\ Lett.\ B {\bf 513}, 232 (2001)
[arXiv:hep-ph/0105239].
%
N.~Arkani-Hamed, A.~G.~Cohen, E.~Katz and A.~E.~Nelson,
arXiv:hep-ph/0206021.
%
N.~Arkani-Hamed, A.~G.~Cohen, E.~Katz, A.~E.~Nelson, T.~Gregoire and J.~G.~Wacker,
arXiv:hep-ph/0206020.

\bibitem{bardeen}
W.~A.~Bardeen,
FERMILAB-CONF-95-391-T
{\it Presented at the 1995 Ontake Summer Institute, Ontake Mountain,
Japan, Aug 27 - Sep 2, 1995}.

\bibitem{fv}
P.~H.~Frampton and C.~Vafa,
arXiv:hep-th/9903226.

\bibitem{malda}
J.~M.~Maldacena,
Adv.\ Theor.\ Math.\ Phys.\  {\bf 2}, 231 (1998)
[Int.\ J.\ Theor.\ Phys.\  {\bf 38}, 1113 (1999)]
[arXiv:hep-th/9711200].

\bibitem{RS}
L.~Randall and R.~Sundrum,
Phys.\ Rev.\ Lett.\  {\bf 83}, 3370 (1999)
[arXiv:hep-ph/9905221].


\bibitem{KRN}
H.~J.~Kim, L.~J.~Romans and P.~van Nieuwenhuizen,
Phys.\ Rev.\ D {\bf 32}, 389 (1985).

\bibitem{arutrecent}
G.~Arutyunov, S.~Penati, A.~C.~Petkou, A.~Santambrogio and E.~Sokatchev,
arXiv:hep-th/0206020.

\bibitem{arut}
G.~Arutyunov and S.~Frolov,
Phys.\ Rev.\ D {\bf 62}, 064016 (2000)
[arXiv:hep-th/0002170].
G.~Arutyunov, S.~Frolov and A.~C.~Petkou,
Nucl.\ Phys.\ B {\bf 586}, 547 (2000)
[Erratum-ibid.\ B {\bf 609}, 539 (2001)]
[arXiv:hep-th/0005182].
E.~D'Hoker, S.~D.~Mathur, A.~Matusis and L.~Rastelli,
Nucl.\ Phys.\ B {\bf 589}, 38 (2000)
[arXiv:hep-th/9911222].

\bibitem{ewdbltr}
E.~Witten,
arXiv:hep-th/0112258.

\bibitem{berkdbltr}
M.~Berkooz, A.~Sever and A.~Shomer,
JHEP {\bf 0205}, 034 (2002)
[arXiv:hep-th/0112264].

\bibitem{absdbltr}
O.~Aharony, M.~Berkooz and E.~Silverstein,
JHEP {\bf 0108}, 006 (2001)
[arXiv:hep-th/0105309].


\bibitem{kleb}
I.~R.~Klebanov and A.~Hashimoto,
Nucl.\ Phys.\ B {\bf 434}, 264 (1995)
[arXiv:hep-th/9409064].

\bibitem{klebnek}
I.~R.~Klebanov and N.~A.~Nekrasov,
Nucl.\ Phys.\ B {\bf 574}, 263 (2000)
[arXiv:hep-th/9911096].
I.~R.~Klebanov and A.~A.~Tseytlin,
Nucl.\ Phys.\ B {\bf 578}, 123 (2000)
[arXiv:hep-th/0002159].

\bibitem{ikms}
I.~R.~Klebanov and M.~J.~Strassler,
JHEP {\bf 0008}, 052 (2000)
[arXiv:hep-th/0007191].

\bibitem{csaki}
C.~Csaki and J.~Terning, private communication.

\bibitem{Troost:2003ig}
J.~Troost,
arXiv:hep-th/0308044.

\bibitem{Gubser:2002zh}
S.~S.~Gubser and I.~Mitra,
Phys.\ Rev.\ D {\bf 67}, 064018 (2003)
[arXiv:hep-th/0210093].

\bibitem{Gubser:2002vv}
S.~S.~Gubser and I.~R.~Klebanov,
Nucl.\ Phys.\ B {\bf 656}, 23 (2003)
[arXiv:hep-th/0212138].

\bibitem{kaplan}
D.~B.~Kaplan,
Phys.\ Lett.\ B {\bf 136}, 162 (1984).


\bibitem{Markus}
H.-S.~Goh, M.A.~Luty and S.-P. Ng,
hep-th/0309103.

\end{references}
\end{document}